\documentclass[prl,reprint,aps,superscriptaddress,onecolumn,nobibnotes,nofootinbib]{revtex4-1}

\usepackage[utf8]{inputenc}
\usepackage{mathptmx}
\usepackage{amsmath}
\usepackage{amssymb}
\usepackage{color,graphicx}
\usepackage[colorlinks,citecolor=blue,urlcolor=blue,linkcolor=blue]{hyperref}
\usepackage[normalem]{ulem}
\usepackage{enumitem}
\usepackage{mathtools}
\usepackage{braket}
\usepackage[per=slash,seperr,tightpm,trapambigerr=false]{siunitx}

\newcommand\Tstrut{\rule{0pt}{2.3ex}}

\date{\today}

\begin{document}
\title{Description of CRESST-III Data}
\newcommand{\mpi}{\affiliation{Max-Planck-Institut f\"ur Physik, 80805 M\"unchen, Germany}}
\newcommand{\coimbra}{\affiliation{Also at: LIBPhys, Departamento de Fisica, Universidade de Coimbra, P3004 516 Coimbra, Portugal}}
\newcommand{\hephy}{\affiliation{Institut f\"ur Hochenergiephysik der \"Osterreichischen Akademie der Wissenschaften, 1050 Wien, Austria}}
\newcommand{\ati}{\affiliation{Atominstitut, Technische Universit\"at Wien, 1020 Wien, Austria}}
\newcommand{\tum}{\affiliation{Physik-Department and Excellence Cluster Universe, Technische Universit\"at M\"unchen, 85747 Garching, Germany}}
\newcommand{\tuebingen}{\affiliation{Eberhard-Karls-Universit\"at T\"ubingen, 72076 T\"ubingen, Germany}} 
\newcommand{\oxford}{\affiliation{Department of Physics, University of Oxford, Oxford OX1 3RH, United Kingdom}}
\newcommand{\wmi}{\affiliation{Also at: Walther-Mei\ss ner-Institut f\"ur Tieftemperaturforschung, 85748 Garching, Germany}}
\newcommand{\lngs}{\affiliation{INFN, Laboratori Nazionali del Gran Sasso, 67010 Assergi, Italy}}
\newcommand{\gssi}{\affiliation{Also at: Gran Sasso Science Institute, 67100, L'Aquila, Italy}}
\newcommand{\cassino}{\affiliation{Also at: Dipartimento di Ingegneria Civile e Meccanica, Universit\'a degli Studi di Cassino e del Lazio Meridionale, 03043 Cassino, Italy}}
\newcommand{\chalmers}{\affiliation{Also at: Chalmers University of Technology, Department of Physics, 412 96 G\"oteborg, Sweden}}

\mpi
\lngs
\tum
\hephy
\ati
\tuebingen
\oxford

\coimbra
\gssi
\wmi
\cassino
\chalmers

\author{A.~H.~Abdelhameed}
  \mpi
  
\author{G.~Angloher}
  \mpi

\author{P.~Bauer}
  \email[Corresponding author: ]{philipp.bauer@mpp.mpg.de}
  \mpi

\author{A.~Bento}
  \mpi
  \coimbra 

\author{E.~Bertoldo}
  \mpi

\author{C.~Bucci}
  \lngs 

\author{L.~Canonica}
  \mpi 

\author{A.~D'Addabbo}
  \lngs
  \gssi

\author{X.~Defay}
  \tum 

\author{S.~Di~Lorenzo}
  \lngs
  \gssi

\author{A.~Erb}
  \tum
  \wmi
  
\author{F.~v.~Feilitzsch}
  \tum 
  
\author{S.~Fichtinger}
  \hephy

\author{N.~Ferreiro~Iachellini}
  \mpi  
  
\author{A.~Fuss}
  \hephy
  \ati

\author{P.~Gorla}
  \lngs 

\author{D.~Hauff}
  \mpi 

\author{J.~Jochum}
  \tuebingen 

\author{A.~Kinast}
  \tum
  
\author{H.~Kluck}
  \hephy
  \ati

\author{H.~Kraus}
  \oxford

\author{A.~Langenk\"amper}
  \tum

\author{M.~Mancuso}
  \mpi
  
\author{V.~Mokina}
  \hephy
  
\author{E.~Mondragon}
  \tum
  
\author{A.~M\"unster}
  \tum

\author{M.~Olmi}
  \lngs
  \gssi
  
\author{T.~Ortmann}
  \tum

\author{C.~Pagliarone}
  \lngs 
  \cassino

\author{L.~Pattavina}
  \tum
  \gssi

\author{F.~Petricca}
  \mpi 

\author{W.~Potzel}
  \tum 

\author{F.~Pr\"obst}
  \mpi

\author{F.~Reindl}
  \hephy
  \ati

\author{J.~Rothe}
  \mpi 
  
\author{K.~Sch\"affner}
  \lngs 
  \gssi

\author{J.~Schieck}
  \hephy
  \ati 

\author{V.~Schipperges}
  \tuebingen
  
\author{D.~Schmiedmayer}
   \hephy
   \ati

\author{S.~Sch\"onert}
  \tum 
  
\author{C.~Schwertner}
  \hephy
  \ati

\author{M.~Stahlberg}
  \hephy
  \ati

\author{L.~Stodolsky}
  \mpi 

\author{C.~Strandhagen}
  \tuebingen

\author{R.~Strauss}
  \tum

\author{C.~T\"urko$\breve{\text{g}}$lu}
  \hephy
  \ati

\author{I.~Usherov}
  \tuebingen 

\author{M.~Willers}
  \tum 

\author{V.~Zema}
  \lngs
  \gssi
  \chalmers

\collaboration{CRESST Collaboration}
\noaffiliation

\begin{abstract}
In CRESST-III, 10 cryogenic detector modules optimized for low energy thresholds were operated for almost two years (May 2016 - February 2018). Together with this document we are publishing data from the best performing detector module which has a nuclear recoil threshold of \SI{30.1}{\electronvolt}. With this data-set we were able to set limits on the cross-section for spin-dependent and spin-independent elastic scattering of dark matter particles off nuclei at dark matter masses down to \SI{160}{\mega\electronvolt\per\clight\squared}. We publish the energies of all events after data selection as well as of all events within the acceptance region for dark-matter searches. In this document we describe how to use these data sets.  
\end{abstract}

\maketitle

\section{Introduction}
\label{sec:intro}

\begin{figure}[htb]
\center
	\includegraphics[width=0.55\textwidth, keepaspectratio]{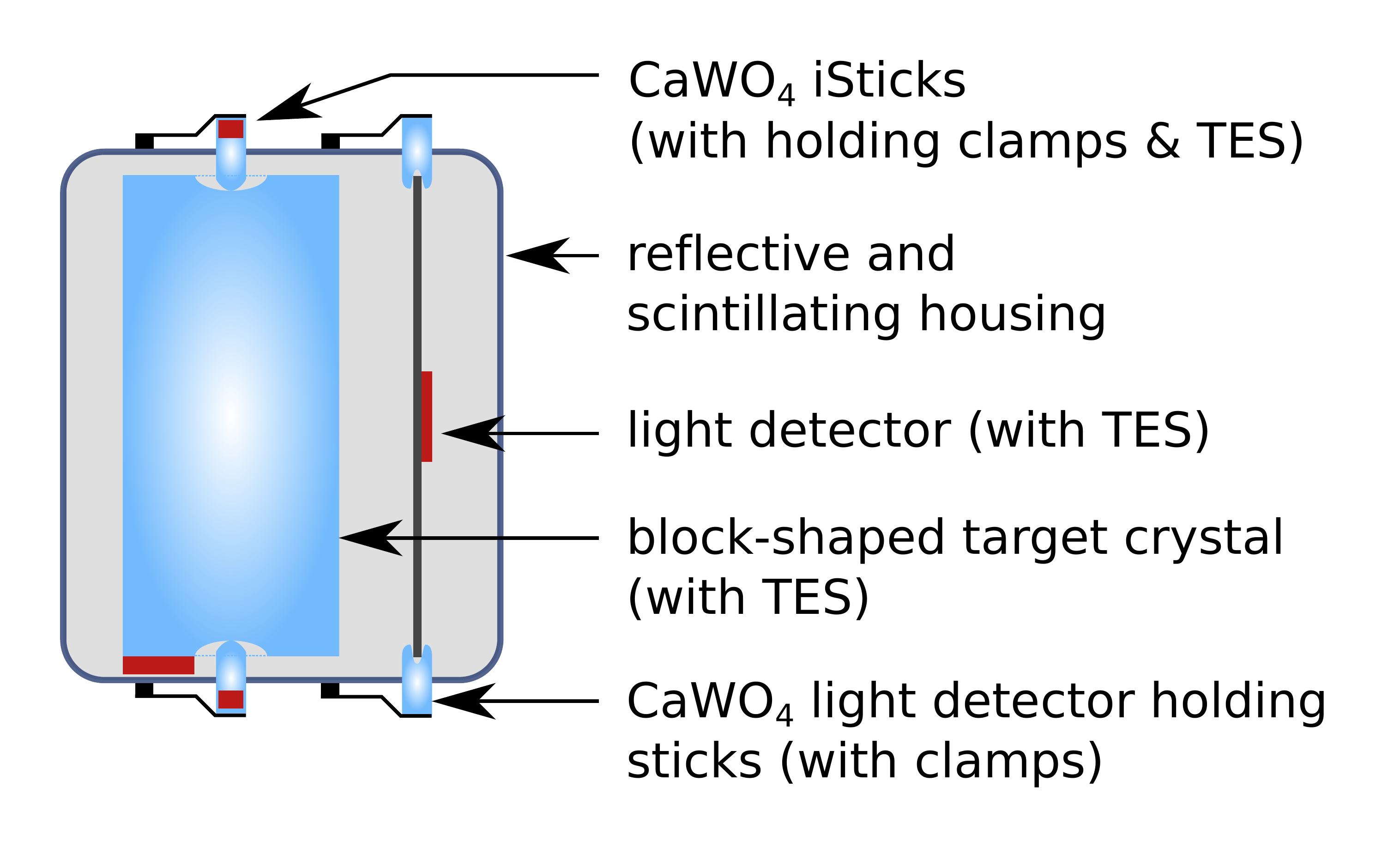}
	\caption{Schematic drawing of the detector module design in CRESST-III}
	\label{fig:Schematics}
\end{figure}

CRESST-III detector modules consist of two cryogenic detectors: The phonon detector, based on a CaWO$_4$ crystal, measures the phonons generated by the energy deposition from an interaction of a particle in the CaWO$_4$ crystal. The light detector, based on a silicon-on-sapphire wafer matched in size to the target crystal, measures the simultaneously created scintillation light. Furthermore, the  CaWO$_4$ sticks holding the phonon detector crystal are also operated as cryogenic detectors (CaWO$_4$ isticks), but serve only as a veto channel for energy depositions in the sticks, see~\cite{strauss_prototype_2017} for further details. A schematic drawing of the detector module design is shown in figure~\ref{fig:Schematics}.

\begin{figure}[h!tb]
	\includegraphics[width=0.48\textwidth]{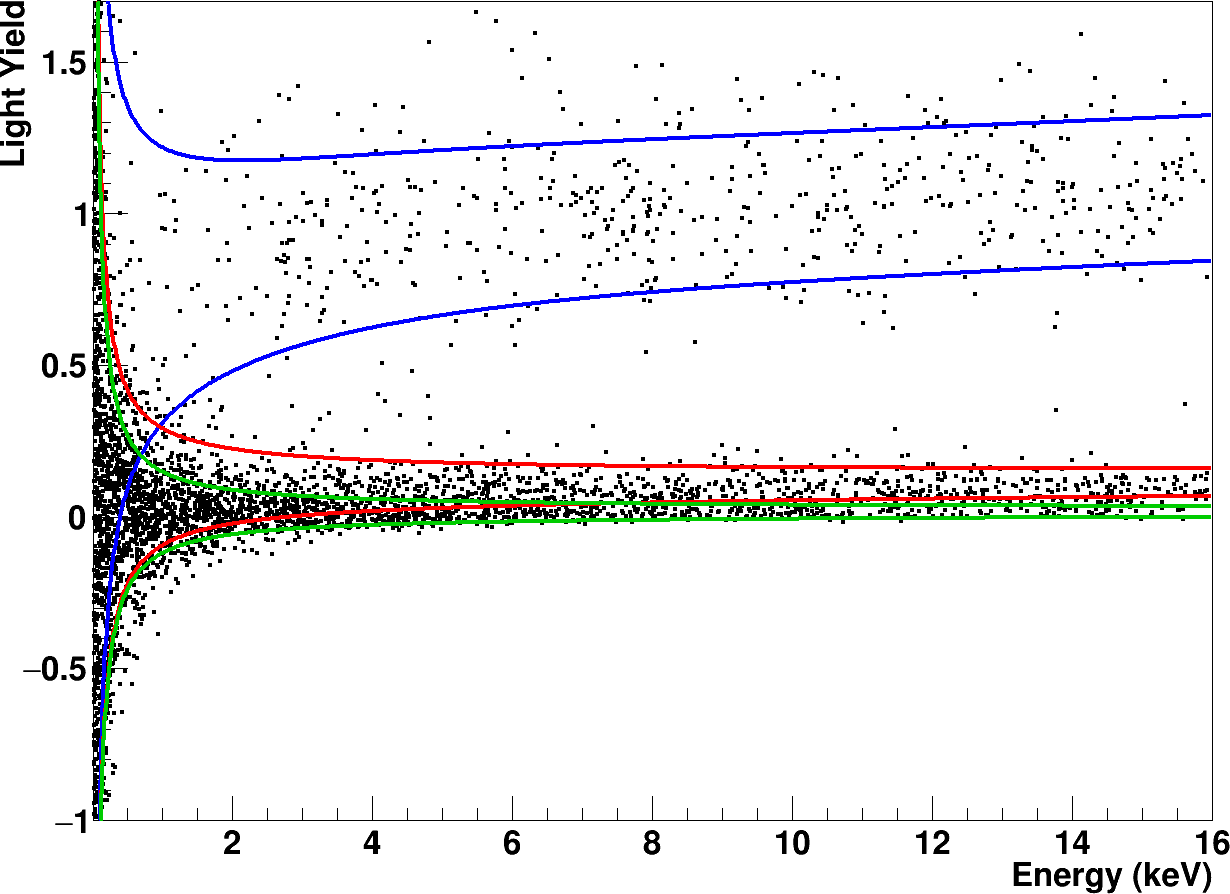}
	\includegraphics[width=0.48\textwidth]{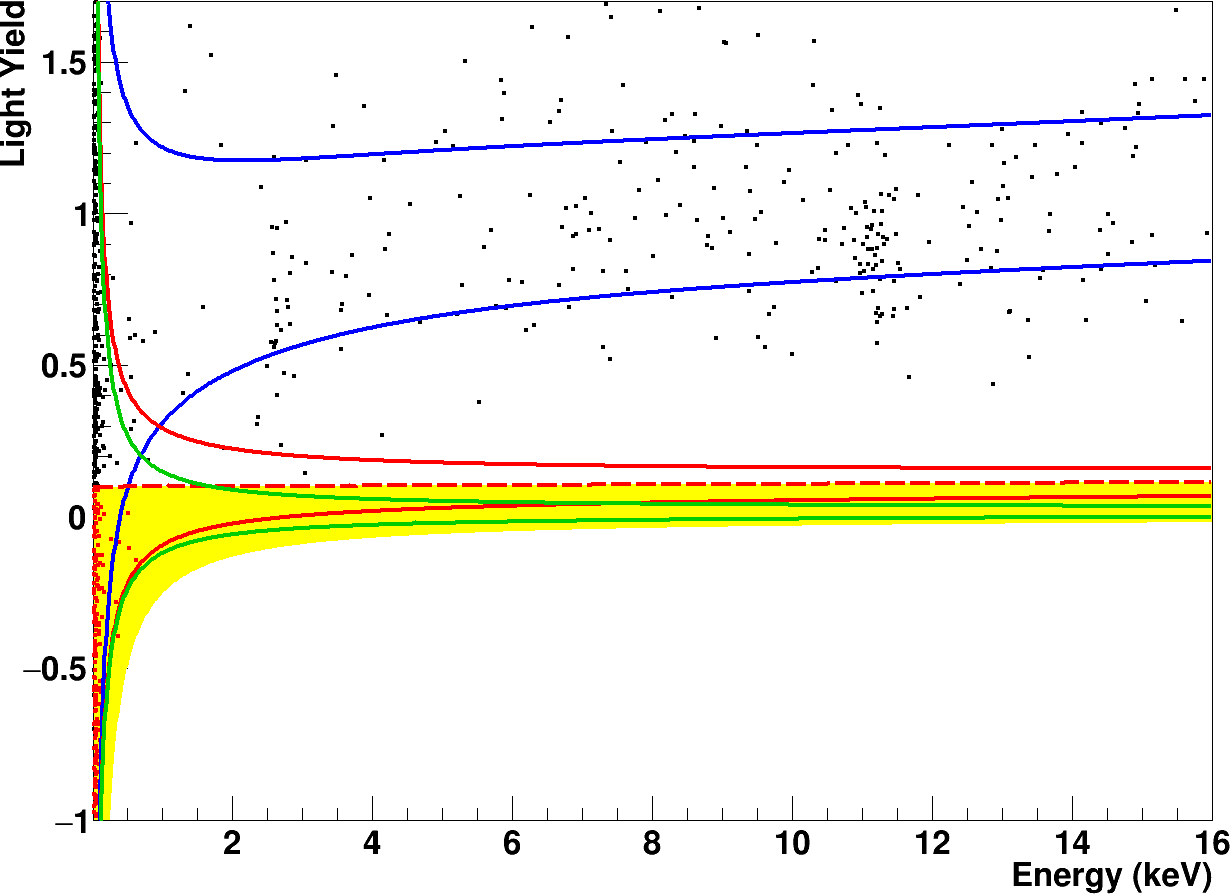}
	\caption{Electron (blue), oxygen (red), and tungsten-recoil bands (green) for detector A. The bands have been fitted to the neutron calibration data (left) and are required for the definition of the acceptance region (yellow) in the blind data (right). The dotted red line is the mean of the oxygen band. The solid lines depict the central \SI{80}{\percent} bands. The $\alpha$ and calcium-recoil bands are not shown for reasons of clarity. Pictures taken from \cite{TheRun34Paper}. An excess of events is observed below \SI{200}{\electronvolt}.}
	\label{fig:Bands}
\end{figure}

Having two read out channels (phonon and light), the light-yield $LY$, i.e., the ratio of the measured light and phonon signals, can be used to distinguish electron from nuclear-recoil events. Dark matter scatters are expected on nuclei, while the dominating backgrounds appear as electron-recoils. The light yield vs energy distribution of events is described by a fitted band description as discussed in \cite{TheRun34Paper} and \cite{Strauss2014} and shown in figure \ref{fig:Bands}. An acceptance region (yellow) is defined for the dark matter search based on this band description. For the CRESST-III analysis in \citep{TheRun34Paper} all events in the acceptance region are regarded as dark matter candidates.

The published data described in this document were obtained from the best performing out of ten detector modules of identical design and were used to set limits on spin-independent as well as spin-dependent dark matter-nucleon interaction cross-sections. The published set only contains events that survive all data selection criteria described in~\cite{TheRun34Paper}. The total exposure before applying any data selection criteria is \SI{5.689}{\kilo\gram\day}. The energy range is restricted to events below \SI{16}{\kilo\electronvolt}.

\section{Comparison of data and models}
For dark matter searches and other studies the energy distribution of measured events is compared to the predictions of different models $p_{\text{model}}(E)$ to obtain or constrain parameters of these models. However, a model $p_{\text{model}}(E)$ has to be corrected for finite energy resolution, energy threshold, and cut-survival probability to be comparable to the energy distribution measured by the detector in reconstructed energy $\widetilde{p}(E_{reco})$:

\begin{equation} \label{equ:DistributionSimple}
	\widetilde{p}(E_{reco}) =  \Theta(E_{reco} - E_{\text{thr,reco}})\cdot\widetilde{\epsilon} \cdot \epsilon_{x,Acc}(E_{reco}) \cdot\int_0^{\infty} p_{\text{model}}(E) \cdot \mathcal{N}(E_{reco}-E, \sigma_p^2)  dE
\end{equation}

$\mathcal{N}(E_{reco}-E, \sigma_p^2)$ is a normal distribution with width $\sigma_p$ being the resolution of the phonon detector for the convolution accounting for the finite energy resolution. $\widetilde{\epsilon}$ is the cut-survival probability that accounts for the loss of signal events from applied data selection criteria. If an acceptance region is defined, the chance of a signal event being in the acceptance region, $\epsilon_{x,Acc}(E_{reco})$, has to be considered as well. The trigger condition imposes a sharp threshold in reconstructed energy, $\Theta(E_{reco} - E_{\text{thr,reco}})$. It is assumed, that the model spectrum $p_{\text{model}}(E)$ is already scaled to the exposure. The required parameters are summarized in table~\ref{tab:threshold}.

\begin{table}[htb]
	\centering
	\begin{tabular}{|l|p{10cm}|}
		\hline
		\textbf{Parameter} & \textbf{Detector A}  \\
		\hline
		\hline
		Exposure before data selection (kgd)& $5.594$\Tstrut\\
		\hline
		$E_{\text{thr,reco}}$\,[keV] & $0.0301 \pm 0.0001$ \Tstrut\\
		\hline
		$\sigma_p$\,[keV] & $0.0046$ \Tstrut \\
		\hline
		$\widetilde{\epsilon}_x$ & $\geq$\SI{50}{\percent} $^1$  \Tstrut\\
		\hline
		$\epsilon_{x,Acc}(E_{reco})$ & see data files \Tstrut\\
		\hline
	\end{tabular}
	\caption{Parameters required for the comparison between data and models. Exposure before data selection, energy threshold $E_{\text{thr,reco}}$ in reconstructed energy, resolution $\sigma_p$ at zero energy and event survival probabilities. }
	\label{tab:threshold}
\end{table}

In general, the energy resolution depends on the energy. The value for $\sigma_p$ given in table~\ref{tab:threshold} is the baseline resolution, i.e., the resolution given by the noise for zero energy deposition. The assumption of constant energy resolution is valid for low energies of $\mathcal{O}$(0.1keV). If required, the energy dependence of the resolution for larger energies can be estimated from lines within the electron-recoil band.

\footnotetext[1]{The detailed cut-survival probability is now given in the appendix \ref{sec:appendix}.} 

\section{Published data-files}

The published data from CRESST-III Phase 1 are stored in ASCII files. The first lines starting with ``\#'' contain a few comments about the contents of each file. The following lines contain the energy of one event per line. 

For the events in the region of interest, files for the survival probabilities $\epsilon_{x,Acc}(E_{reco})$ for nuclear recoils on tungsten, calcium and oxygen are also made available. Similar to the event-data files, the first lines starting with ``\#'' contain a few comments about the contents of each file. The following lines contain $\epsilon_{x,Acc}(E_{reco}))$ arranged in two columns:
\begin{enumerate}
  \item The energy $E_{reco}$ in keV.
  \item $\epsilon_{x,Acc}(E_{reco})$ 
\end{enumerate}

\subsection{Data after event selection}

\begin{figure}[h!tb]
\center
	\includegraphics[width=0.48\textwidth]{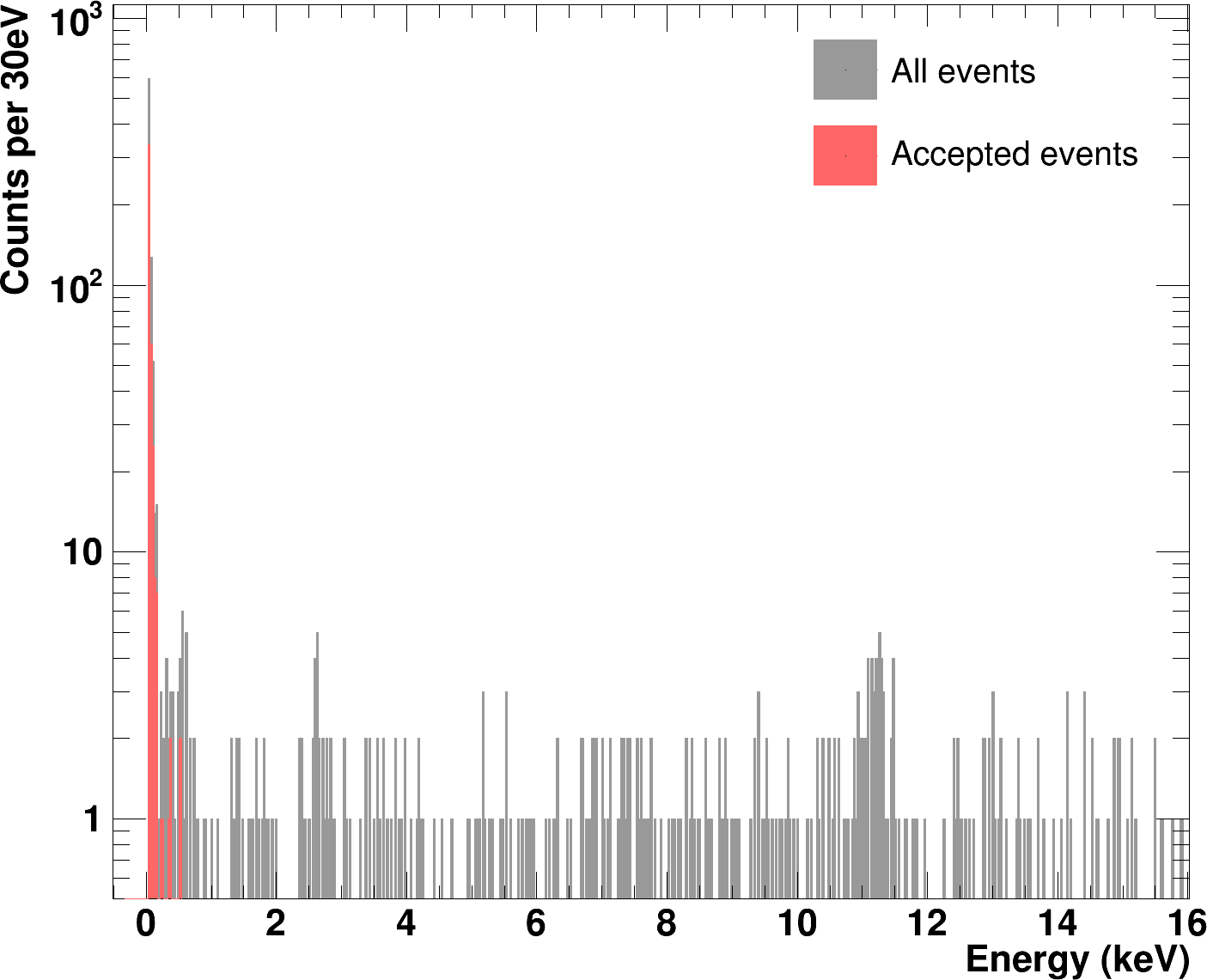}
	\caption{Energy spectrum for events in the full data set (gray) and the acceptance region (red). Figure taken from \cite{TheRun34Paper}.}
	\label{fig:EnergySpectrum}
\end{figure}

The file \verb=C3P1_DetA_full.dat= contains the energies of all events after data selection but without consideration of the acceptance region i.e. $\epsilon_{x,Acc}(E_{reco})=1$. Above \SI{3}{\kilo\electronvolt} there are no nuclear recoils observed (compare figure~\ref{fig:Bands}). The energy spectrum is shown in gray in figure~\ref{fig:EnergySpectrum}.

\subsection{Acceptance region for dark-matter search}
\label{sec:AcceptanceData}
\begin{figure}[h!tb]
\center
	\includegraphics[width=0.55\textwidth]{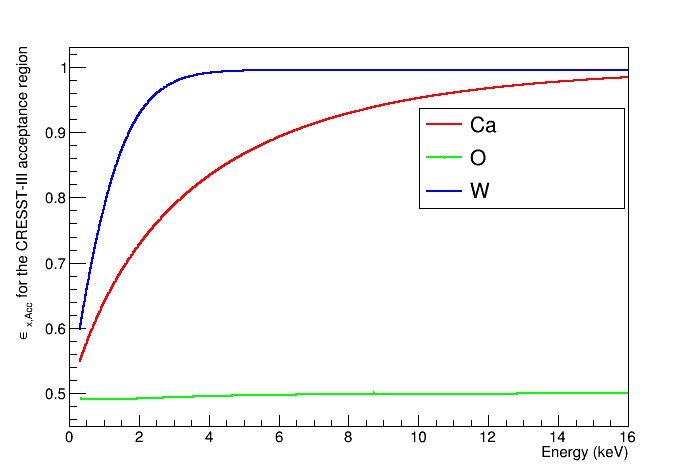}
	\caption{$\epsilon_{x,Acc}(E_{reco})$ for the acceptance region of the dark matter search of CRESST-III for nuclear recoils on oxygen, calcium and tungsten.}
	\label{fig:RoiLightYieldCutEff}
\end{figure}

The file \verb=C3P1_DetA_AR.dat= contains the energies of events within the acceptance region depicted in yellow in figure~\ref{fig:Bands}. It extends from the mean of the oxygen band down to the lower \SI{99.5}{\percent} boundary of the tungsten band. Details concerning the choice of acceptance region can be found in \cite{TheRun34Paper}. The corresponding \verb=C3P1_DetA_eff_AR_X.dat= files contain the different $\epsilon_{x,Acc}(E_{reco})$s shown in figure~\ref{fig:RoiLightYieldCutEff} for calcium recoils (red), oxygen recoils (green) and tungsten recoils (blue). Only events within the interval [\SI{30.1}{\electronvolt},\SI{16}{\kilo\electronvolt}] are accepted. Figure~\ref{fig:EnergySpectrum} shows the energy spectrum of the accpted events in red. See \cite{TheRun34Paper}, for a discussion of the features of the spectrum.

\section{Citation}

If you base your work on our data, we kindly ask you to cite this document as well as \cite{TheRun34Paper}.

\bibliographystyle{h-physrev}
\bibliography{DataDescription.bib}

\section{Appendix} \label{sec:appendix}
This appendix was added on April 04, 2020 by F. Reindl (\href{mailto:florian.reindl@oeaw.ac.at}{florian.reindl@oeaw.ac.at}). No modifications to the content above were done.

\subsection{Limits calculated with analytic convolution}
As discussed in section IV.5.~of \cite{TheRun34Paper} the dark matter expectation (the expected recoil energy spectrum) was obtained using simulated events superimposed to measured noise. This method was chosen as it automatically accounts for all effects and efficiencies of triggering, energy reconstruction and event selection. 

\begin{figure}[h!tb]
	\includegraphics[width=0.48\textwidth]{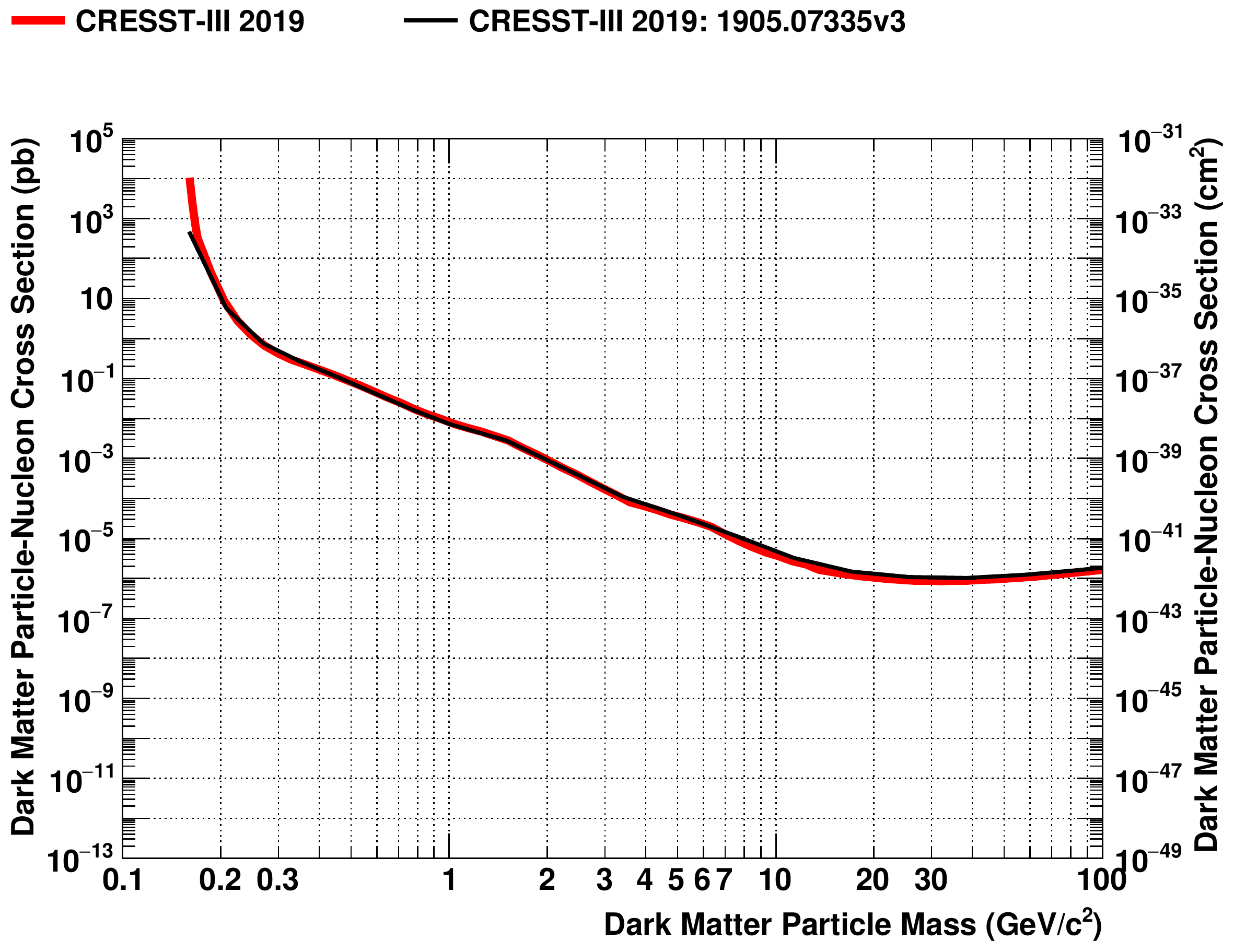}
	\includegraphics[width=0.48\textwidth]{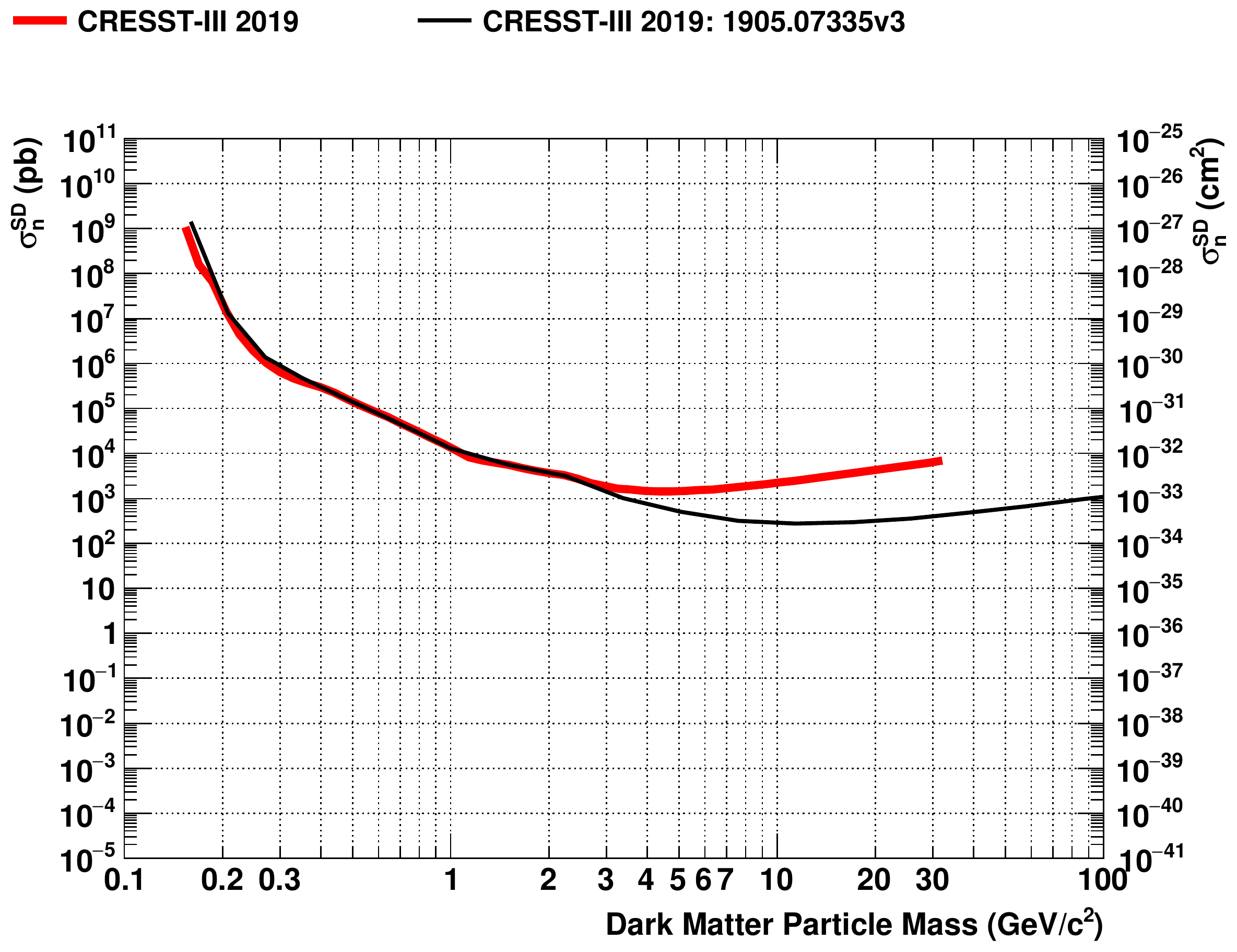}
	\caption{Comparison of limits for spin-independent interactions (left) and spin-dependent interactions (right). Red: as published in \cite{TheRun34Paper} using a simulation for the expected dark matter recoil energy spectrum; black: using an analytic calculation for the expected dark matter recoil energy spectrum following the receipt of this data release. Limits in files C3P1\_DetA\_DataRelease\_SI/SD.xy.  }
	\label{fig:Limits}
\end{figure}

However, for the paper at hand we suggested to use an analytic calculation of the expected recoil spectrum where the finite detector resolution is taken into account via a convolution with a Gaussian function, see equation \ref{equ:DistributionSimple}. We find very little difference for both methods in the spin-independent case, while for the spin-dependent case and higher masses the method described in \cite{TheRun34Paper} leads to more conservative limits. To allow for best reproducibility with this paper we are publishing new limits in this release calculated as outlined above. The corresponding files are named \verb=C3P1_DetA_DataRelease_SI/SD.xy=. Figure \ref{fig:Limits} depicts a comparison of these limits (black) to the limits published in \cite{TheRun34Paper} (red). 

\subsection{Energy-dependent cut-survival probability $\widetilde{\epsilon}_x$}

Using this energy-dependent efficiency instead of a simple constant factor of 50~\% improves the accuracy of the limit for very low dark matter particle masses, for both the spin-independent and spin-dependent case. 

\begin{figure}
  \centering
  \includegraphics[width=0.5\textwidth]{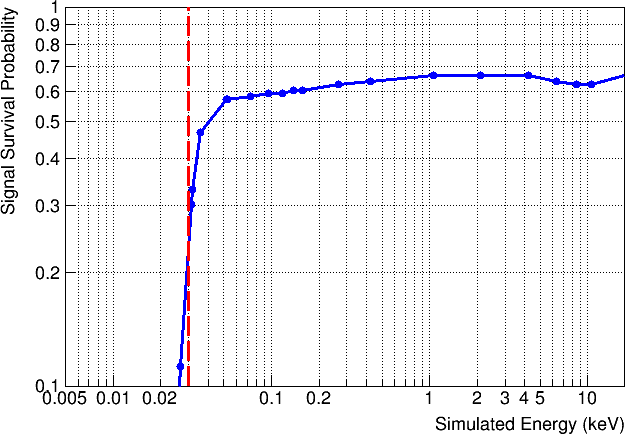}
  \caption{Cut-survival probability $\widetilde{\epsilon}_x$ as a function of simulated energy. Plot of file C3P1\_DetA\_cuteff.dat.}
  \label{fig:cuteff}
\end{figure}

In table \ref{tab:threshold} we give an approximate value cut-survival probability $\widetilde{\epsilon}_x$ (50~\%), while in figure \ref{fig:cuteff} (\verb=C3P1_DetA_cuteff.dat=) we now also provide the precise energy-dependent behavior for $\widetilde{\epsilon}_x$. Using this energy-dependent efficiency in particular improves the accuracy of the limit for very low dark matter particle masses, for both the spin-independent and spin-dependent case. 

\end{document}